\documentclass[sigconf,natbib=true]{acmart}

\AtBeginDocument{%
  \providecommand\BibTeX{{%
    \normalfont B\kern-0.5em{\scshape i\kern-0.25em b}\kern-0.8em\TeX}}}
    
\usepackage{subcaption}
\usepackage{graphicx}
\usepackage{caption}
\usepackage{amsmath}
\usepackage{titlesec}
\usepackage{pifont}
\usepackage[]{xcolor}

\usepackage{algorithm}
\usepackage{algpseudocode}
\algrenewcommand\algorithmicrequire{\textbf{Input:}}
\algrenewcommand\algorithmicensure{\textbf{Output:}}

\usepackage{algpseudocode}
\usepackage{threeparttable}
\usepackage{multirow}
\usepackage{tabularx}
\usepackage{cleveref}
\usepackage{enumitem}

\titlespacing*{\section}{0pt}{1pt plus 0pt}{1pt minus 1pt}
\titlespacing*{\subsection}{0pt}{1pt plus 1pt}{1pt minus 1pt}
\titlespacing*{\subsubsection}{0pt}{1pt}{1pt minus 1pt}
\copyrightyear{2023}
\acmYear{2023}
\setcopyright{rightsretained}
\acmConference[SIGIR '23]{Proceedings of the 46th International ACM SIGIR
Conference on Research and Development in Information Retrieval}{July 23--27,
2023}{Taipei, Taiwan}
\acmBooktitle{Proceedings of the 46th International ACM SIGIR Conference on
Research and Development in Information Retrieval (SIGIR '23), July 23--27,
2023, Taipei, Taiwan}\acmDOI{10.1145/3539618.3591936}
\acmISBN{978-1-4503-9408-6/23/07}


\begin{document}


\title{A Lightweight Constrained Generation Alternative for Query-focused Summarization}

\author{Zhichao Xu}
\email{zhichao.xu@utah.edu}
\affiliation{%
  \institution{University of Utah}
  \city{Salt Lake City}
  \state{Utah}
  \country{United States}
  }
\author{Daniel Cohen}
\email{daniel.cohen@dataminr.com}
\affiliation{%
  \institution{Dataminr, Inc.}
  \city{NYC}
  \state{NY}
  \country{United States}
  }

\begin{abstract}
Query-focused summarization (QFS) aims to provide a summary of a document that satisfies information need of a given query and is useful in various IR applications, such as abstractive snippet generation.
Current QFS approaches typically involve injecting additional information, e.g. query-answer relevance or fine-grained token-level interaction between a query and document, into a finetuned large language model. 
However, these approaches often require extra parameters \& training , and generalize poorly to new dataset distributions. 
To mitigate this, we propose leveraging a recently developed constrained generation model Neurological Decoding (NLD) as an alternative to current QFS regimes which rely on additional sub-architectures and training. 
We first construct lexical constraints by identifying important tokens from the document using a lightweight gradient attribution model, then subsequently force the generated summary to satisfy these constraints by directly manipulating the final vocabulary likelihood. This lightweight approach requires no additional parameters or finetuning as it utilizes both an off-the-shelf neural retrieval model to construct the constraints and a standard generative language model to produce the QFS.
We demonstrate the efficacy of this approach on two public QFS collections achieving near parity with the state-of-the-art model with substantially reduced complexity.

\end{abstract}

\maketitle

\vspace{-5pt}
\section{Introduction}
\label{sec:intro}

In modern search systems, users are often presented with short \textit{snippets} of a candidate document on their search results page.
This snippet serves as a critical element in helping users determine whether a document satisfies their information needs without requiring them to invest additional time. The effectiveness of a snippet largely depends on its ability to accurately and concisely capture the relevant information from the corresponding document in just a few lines of text~\cite{park2022qsg,dang2006duc}.

This task of query-focused summarization (QFS) snippet generation, commonly referred to as query-biased summarization~\cite{tombros1998advantages} or abstractive snippet generation~\cite{chen2020abstractive}, aims to construct a summary that succinctly addresses the information need of a query by extracting essential information from a document. Traditionally, QFS has used extractive methods that rely on the most relevant spans of text from a candidate document based on the prevalence of query terms~\cite{bast2014efficient, turpin2007fast}. Although efficient, this extractive approach is constrained by the format of the original document, with the effectiveness of the snippet heavily dependent on the length and information density of the candidate document \cite{tombros1998advantages}. Moreover, this paradigm limits the possibility of personalization or multiple-document QFS.

With the advent of large language models (LM), a new paradigm has emerged that attempts to address the limitations of extractive snippet generation. These LM-based approaches directly generate abstractive snippets that do not necessarily appear anywhere in the original document~\cite{chen2020abstractive, park2022qsg, lewis2019bart}. While these methods hold promise, successful application is nontrivial. They often require specific architectures with additional parameters to incorporate the relevance signal, along with extensive fine-tuning. Moreover, a significant challenge with natural language generation, including QFS, is the problem of hallucination, where the model confidently generates false information~\cite{maynez2020faithfulness,ji2022survey}. This can lead to unreliable snippets that do not accurately reflect the content of the original document. 


In this paper, we propose a novel lightweight alternative approach for QFS, relevance-constrained QFS, that achieves near parity with current state-of-the-art methods with significantly reduced complexity. Our approach does not require training an additional model or adding more parameters to the existing one. Instead, we demonstrate that QFS can be effectively and efficiently achieved by constraining a general language model (LM) to summarize the document with predefined lexical constraints. We hypothesize that QFS can be effectively achieved by enforcing the language model to summarize with predefined lexical constraints -- i.e, constraining a pretrained LM to favor the most important terms for the relevance of the document. While the notion of constrained generation  \cite{lu2020neurologic} has emerged as a way to combat hallucination in other natural language generation tasks such as commonsense generation, constrained machine translation, conversation \cite{wang2023depth} and table to text generation \cite{lu2021neurologic}, we show the unique advantages it possesses in the case of abstractive QFS.

To achieve this, we first identify the most critical tokens from the ranking model using the gradient signal of each token~\cite{sanchez2015towards} as these salient terms capture the most important aspects of the document's relevance to the query. 
We then convert these tokens into predicate logic constraints and use them as input to a version of constrained generation, Neurological Decoding~\cite{lu2020neurologic}.
By constraining the LM to simultaneously satisfy these constraints and maintain fluency, we generate an abstractive summary that is optimized for relevance to the query. 
This approach allows us to effectively generate snippets without requiring additional complex modules or training methods, making it a lightweight yet effective alternative to the current state-of-the-art method.

Our experiments on two benchmark snippet generation datasets \cite{nema2017diversity,jin2019pubmedqa} demonstrate that this application of relevance-constrained QFS achieves comparable results to the current state-of-the-art method, suggesting a promising alternative perspective to the snippet generation task.


\section{Related Work}
\label{sec:related}

\vspace{0pt}
\textbf{\textit{Query-focused Summarization:}} \space
To generate a query-focused summary, several studies used an additional query-attention mechanism. 
QR-BERTSUM-TL \cite{laskar2020query} incorporates query relevance scores into  a pre-trained summarization model.
\citet{su2021improve} propose merging the representation of an answer span predicted by a separate QA model into the Seq2Seq model's training and inference process to enforce the summary's coherence w.r.t. the query.
QSG Transformer \cite{park2022qsg} suggests using a separate graph neural network model to learn per-token representations and fuse them to the Seq2Seq model to effectively generate a QFS.
These mechanisms can be viewed as enforcing soft semantic constraints during the generation process, and requires additional modules and parameters to function effectively.
We opt for a different approach, i.e. explicitly enforcing lexical constraints during the generation process, without the additional machinery that is necessary to handle the soft semantic constrains.

\noindent
\textbf{\textit{Constrained Generation}} \space
(or Conditional Generation) is a family of natural language generation (NLG) methods that aim to generate natural language including/excluding a set of specific words, i.e. lexical constraints.
The NLG domain recipe leverages pre-trained large language models (LLM) finetuned on specific datasets \cite{erdem2022neural}. 
However, as pointed out by~\citet{lu2020neurologic}, such models only finetuned in an end-to-end manner do not learn to follow the underlying constraints reliably even when supervised with large amounts of training examples.
Therefore, a line of works \cite{lu2020neurologic,lu2021neurologic,anderson2016guided,hokamp2017lexically} in constrained generation proposes to explicitly modify the likelihood of next word prediction in the generation stage, such that the predefined lexical constraints can be better satisfied.

\section{Relevance-Constrained QFS}
\label{sec:method}
\vspace{0pt}
\noindent
\textbf{\textit{Problem Formulation:}} \space 
Given a query-document pair $(q,d)$, our task is to generate an abstract summarization $s$, which addresses the information need of the query. 

We propose addressing this problem by leveraging a relevance-constrained generation. In this section, we first introduce how we construct the set of constraints used by the language model to generate the abstract summary. We then present the constrained generation process itself.


\vspace{3pt}
\noindent
\textbf{\textit{Identifying Constraints:}} \space 
In order to identify the most effective constraints for QFS, we first assume that each candidate document is relevant to the query. We then use a pointwise cross-entropy loss, $\mathcal{L}$, to identify how each token contributes to the relevance of the document. To achieve this, we use a saliency based mapping approach to quantify this impact as gradient-based attribution methods have been widely adopted in existing NLP literature \cite{feng2018pathologies,ross2020explaining,wang2020gradient}.

Formally, denote an input sequence $(w_1, w_2, \cdots, w_n)$, where $w_i$ is the $i$-th token; and $\mathbf{x}=(\mathbf{x}_1, \mathbf{x}_2, \cdots, \mathbf{x}_n)$ is a sequence of corresponding static token embeddings.
Let $f(\cdot)$ be a function that takes $\mathbf{x}$ as input and outputs a prediction logit, e.g., a transformer-style model with classification head.
The gradients w.r.t. each input token $w_i$ can be regarded as each token's contribution, or \textit{saliency}, to the final prediction $f(\mathbf{x})$. 
We denote this per token gradient vector as $\mathbf{a}=(a_1, a_2, \cdots, a_n)$, which is the normalized saliency across all tokens,
\begin{equation}
    \vspace{-2pt}
    a_i = \frac{g(\nabla_{\mathbf{x}_i} \mathcal{L},\,\mathbf{x}_i)}{\sum_{j=1}^n  g(\nabla_{\mathbf{x}_j} \mathcal{L}, \,\mathbf{x}_j) }
    \label{eq:attribution}
    \vspace{-2pt}
\end{equation}
where $\mathcal{L}$ denotes the loss between $f(\mathbf{x})$ and label $y=1$, and $g(\cdot,\cdot)$ is the saliency function.

While there exists various methods to estimate the saliency via $g(\cdot, \cdot)$ \cite{sundararajan2017axiomatic,simonyan2013deep,smilkov2017smoothgrad, feng2018pathologies, wang2020gradient}, we adopt \texttt{InteGrad} \cite{sundararajan2017axiomatic}, as it is robust to input perturbations~\cite{wang2020gradient}. 
Specifically, \texttt{InteGrad} sums the gradients along the path from a baseline input $\mathbf{x}_i' = \mathbf{0}$ to the actual input $\mathbf{x}_i$:
\begin{equation}
    \vspace{-2pt}
    g(\nabla_{\mathbf{x}_i} \mathcal{L}, \mathbf{x}_i) = (\mathbf{x}_i -\mathbf{x}_i') \times \sum_{k=1}^m \frac{\partial f(\mathbf{x}_i' + \frac{k}{m} \times (\mathbf{x}_i -\mathbf{x}_i'))}{\partial \: \mathbf{x}_i}
    \label{eq:integrad}
    \vspace{-2pt}
\end{equation}
where $m$ is the number of steps to interpolate the input $\mathbf{x}_i$ and $\times$ denotes dot product; thus $g(\nabla_{\mathbf{x}_i} \mathcal{L}, \mathbf{x}_i)$ is a scalar indicating saliency of token $w_i$ before normalization (Eq.~1). In our implementation, we follow the original setup in \cite{sundararajan2017axiomatic} and set $m$ to 10 steps. 

We note that any differentiable retrieval function can be used in place of $f(\cdot)$ within this framework. In this paper, we use a standard DistilBERT document reranker trained on MS MARCO using a cross-entropy loss~\cite{nogueira2020document, gao2021rethink, boytsov2022understanding, nguyen2016ms}.  

In our preliminary experiments, we observed that the saliency scores are often noisy, attributing gradients to stopwords and/or punctuations.
Therefore, we filter out the stopwords and punctuations in a post hoc manner and only keep the top-3 important tokens from document $d$ to construct the actual decoding constraints $\mathcal{C}$.

\noindent
\textbf{\textit{Constructing Constraints:}} \space Having identified the most salient tokens, we construct the lexical constraints in a format appropriate for constrained generation, Conjunctive Normal Form,
\vspace{-2pt}
$$
\mathcal{C} = \underbrace{(D_1 \vee D_2 \vee \cdots \vee D_i)}_{C_1} \wedge \cdots \wedge \underbrace{(D_k \vee D_{k+1} \vee \cdots \vee D_n )}_{C_m}
$$
where each single $D_i$ denotes one single positive or negative constraint, which we refer to as a \emph{literal}; and the logical disjunction of literals is referred to as a \emph{clause}, e.g. $C_1$ to $C_m$. 
In our implementation, we construct 3 clauses with each clause initially consisting of a single literal. We then expand each clause by all possible forms of the original token via WordForms\footnote{https://github.com/gutfeeling/word\_forms}.
An example of this logic corresponding to Row~1, Table~\ref{tab:qualitative} is represented as
\begin{align*}
\mathcal{C} =& \underbrace{(\text{private} \vee \ldots \vee \text{privatization})}_{C_1} \wedge \underbrace{(\text{health} \vee \ldots \vee \text{healthy})}_{C_2} \\ 
&\wedge \underbrace{(\text{standard} \vee \ldots \vee \text{standards})}_{C_3}
\end{align*}

\vspace{-2pt}
\noindent
\textbf{\textit{Constrained Generation:}} \space 
At inference time, we run a simplified version of the Neurological Decoding (NLD) algorithm using the set of constraints $\mathcal{C}$ acquired from Section~\ref{sec:method}. 
As we do not use negative constraints in QFS, i.e. we do not avoid certain tokens, we consider only two states within the original NLD algorithm: \textit{reversible unsatisfaction} where an unsatisfied logical clause with a positive literal can be satisfied at a future point and \textit{irreversible satisfaction} where a positive literal will remain satisfied. 
This predicate logic is then applied within a conventional beam search during generation. 

At timestep $t$, the simplified algorithm performs three individual steps when filling in beam candidates: \emph{Pruning}, \emph {Grouping}, and \emph{Selecting}. 
Pruning filters out candidates that are of low likelihood or satisfy fewer clauses; Grouping implicitly constructs the power set of all irreversible satisfied clauses, leading to at most $2^{|\mathcal{C}|}$ groups; and Selecting populates the beam with candidates within each group that are most likely to satisfy remaining reversible unsatisfied clause $C_j$ by modifying the likelihood. 
Specifically, within each group, the likelihood is modified by the NLD score function:
\vspace{-2pt}
\begin{equation}
    L = P_\theta (y_t| y_{<t}) + \lambda \max_{\mathbb{I}(C_j)=0} \frac{|\hat{D}_i|}{|D_i|}
    \label{eq:prune}
\end{equation}
where $P_\theta$ is the likelihood of the LM generating token $y_t$, $\mathbb{I}(C_j)$ indicates whether clause $C_j$ has been satisfied or not, 
 $\frac{|\hat{D_i}|}{|D_i|}$ is the overlap between the ongoing generation and the partially satisfied literal $D_i$, e.g. $\hat{D_i}=$"apple" and ${D_i}=$"apple tree" yields 0.5, and $\lambda=0.1$ acts as the hyperparameter. Intuitively, this score modification favors candidates moving toward fully satisfying a positive literal within an unsatisfied clause with $\lambda$ controlling the strength of this signal.
After this explicit likelihood modification, we visit each group and select the highest scoring candidate in rotation until the beam is filled.
After this process is complete, we select the beam candidate with highest score and proceed to generating the next token at $t+1$.
Although the group construction suggests a high-complexity runtime, implicit construction results in this algorithm having the same runtime complexity as standard beam search~\cite{lu2020neurologic}.


We use BART \cite{lewis2019bart} and T5 \cite{raffel2020exploring} for fair comparison with existing methods as the generating LM for abstractive QFS. As there exist no additional parameters or modules for this method, details of these backbone LMs are discussed in Section~\ref{sec:experiment}.

\section{Experimental Setup}
\label{sec:experiment}

\begin{table}[t]
\centering
\caption{Results on test set, including ROUGE-1, ROUGE-2 and ROUGE-L, baseline results (the first section) are from \cite{park2022qsg}; \emph{Italic} indicates the best performing system in literature. $\dag$ denotes the constrained method significantly better than its unconstrained counterparts with paired t-test at 0.05 level}
\resizebox{\columnwidth}{!}{
\begin{tabular}{lccccccc}
\toprule
 \multirow{2}{*}{Model} & \multicolumn{3}{c}{\texttt{Debatepedia}} & & \multicolumn{3}{c}{\texttt{PubMedQA}}  \\ \cline{2-4} \cline{6-8}
 & R-1 & R-2 & R-L &  & R-1 & R-2 & R-L \\ \midrule
    Transformer & 41.7 & 33.6 & 41.3 &  & 30.4 & 8.4 & 22.3 \\
    SD2  & 41.3 & 18.8 & 40.4 &  & 32.3 & 10.5 & 26.0 \\
    CSA Transformer & 46.4 & 37.5 & 45.9 &  & - & - & - \\
    QR-BERTSUM-TL & 48.0 & 45.2 & 57.1 &  & - & - & - \\
    MSG & - & - & - &  &  37.2 & 14.8 & $\textbf{30.2}$ \\
    BART-QFS & 59.0 & 44.6 & 57.4 &  & - & - & -  \\
    \emph{QSG BART} & $\textbf{64.9}$ & $\textbf{52.3}$ & $\textbf{63.3}$ &  & $\textbf{38.4}$ & $\textbf{17.0}$ & $29.8$ \\ \midrule
    T5 & 22.5 & 7.1 & 19.7 &  & 38.0 & 15.3 & 28.2 \\
    Constrained-T5 & $32.2^\dag$ & $12.5^\dag$ & $28.2^\dag$ &  & 36.4 & $16.0^\dag$ & $28.7^\dag$ \\
    - Rel. Improv. (\%) & $+43.1$ & $+76.1$ & $+43.2$ &  & $-4.3$ & $+4.6$ & $+1.8$ \\
    BART & 58.1 & 43.6 & 56.8 &  & 38.1 & 15.7 & 27.2 \\
    Constrained-BART & $\textbf{62.9}^\dag$ & $\textbf{50.1}^\dag$ & $\textbf{61.5}\dag$ &  & $\textbf{39.2}^\dag$ & $\textbf{17.1}^\dag$ & $\textbf{30.1}^\dag$ \\
    - Rel. Improv. (\%) & $+8.3$ & $+14.9$ & $+8.3$ &  & $+2.9$ & $+8.9$ & $+10.6$ \\
    \bottomrule
    \end{tabular}
}
    \label{tab:results}
    \vspace{-2pt}
\end{table}

\vspace{0pt}
\noindent
\textbf{\textit{Datasets:}} \space
Following previous works \cite{su2021improve,park2022qsg}, we adopt Debatepedia \cite{nema2017diversity} and PubMedQA \cite{jin2019pubmedqa} to benchmark the effectiveness of the proposed relevance-constrained generation method.
Debatepedia dataset is collected by \citet{nema2017diversity} from 663 debates of 53 diverse categories in an encyclopedia of debates and consists of 12K/0.7K/1.0K query-document summarization triplets $(q,d,s)$.
PubMedQA is a long-form abstractive question-answering dataset from the biomedical domain with the contexts available. We use the standard train test split from the original datasets. 

\noindent
\textbf{\textit{Compared Methods:}} \space
To evaluate the performance of the proposed relevance-constrained generation method, we introduce the following baseline methods in order of increasing complexity: \\
\vspace{1pt}
\textbullet \, \textbf{End-to-End approaches}: Transformer \cite{vaswani2017attention}, BART \cite{lewis2019bart} and T5 \cite{raffel2020exploring} are finetuned for Seq2Seq summarization.
    These LMs additionally act as the backbone LM for the proposed relevance-constrained QFS approach, i.e. Constrained-BART and Constrained-T5 such that the results are directly comparable. In this configuration, there are no constraints during the generation process. \\
\vspace{1pt}
\textbullet \, \textbf{Improved query-document cross attention}: SD2 \cite{nema2017diversity} adds additional cross attention between query and document encoder, then uses the combined representation for generation. CSA Transformer \cite{xie2020conditional} adds conditional self-attention layers originally designed for conditional dependency modeling to the Seq2Seq model.  \\
\vspace{1pt}
\textbullet \, \textbf{Incorporated query-document relevance}: QR-BERTSUM-TL \cite{laskar2020query} injects query relevance scores into pretrained Seq2Seq summarization model; 
MSG \cite{deng2020multi} utilizes query relevance and interrelation between sentences of the document for fine-grained representation.
Similarly, BART-QFS \cite{su2021improve} also uses a pre-trained QA model to determine answer relevance in the document and injects this information into the Seq2Seq LM model. \\
\vspace{1pt}
\textbullet \, \textbf{Additional module utilization}: QSG-BART \cite{park2022qsg} utilizes an additional graph neural network module to model token-level interaction between query and document, and injects this information into Seq2Seq model. It reaches state-of-the-art performance on the QFS task, but requires additional parameters and training.

\begin{table*}[t]
    \centering
    \caption{Sample qualitative study on Debatepedia dataset; \colorbox{SkyBlue}{tokens} are marked salient and included in constraints set $\mathcal{C}$.}
    \resizebox{1.0\textwidth}{!}{
    \begin{tabular}{l|l|l|l|l}
    \toprule
    \multicolumn{1}{c|}{Query} & \multicolumn{1}{c|}{Document} & \multicolumn{1}{c|}{BART Generation} & \multicolumn{1}{c|}{Constrained-BART Generation} & \multicolumn{1}{c}{Golden Reference} \\ \hline
    \begin{tabular}[c]{@{}l@{}l@{}l@{}l@{}l@{}l} privatization: is water\\ a resource that should\\ be owned by private\\ companies versus a\\ global commons?  \end{tabular} & 
    \begin{tabular}[c]{@{}l@{}l@{}l@{}l@{}l@{}l@{}l} \colorbox{SkyBlue}{\textbf{private}} companies are profit-maximizing entities\\that often view environmental \colorbox{SkyBlue}{\textbf{health}} and safety\\ \colorbox{SkyBlue}{\textbf{standards}} as obstructive to their profit interests.\\ this is a problem particularly in the context of water\\ which is so fundamentally important to the environ-\\ment health and life. \end{tabular} & 
    \begin{tabular}[c]{@{}l@{}l@{}l} environmental and health\\ standards are often violated\\ by public owners of water\\ \end{tabular} &
    \begin{tabular}[c]{@{}l@{}l@{}l@{}l} environmental and \colorbox{SkyBlue}{\textbf{health}}\\ \colorbox{SkyBlue}{\textbf{standards}} are often violated\\ by water \colorbox{SkyBlue}{\textbf{privatization}}\\ companies.\end{tabular} &
    \begin{tabular}[c]{@{}l@{}l@{}l} environmental and health\\ standards are often violated\\ by private ownership of water\\ \end{tabular}
    \\ \hline
    \begin{tabular}[c]{@{}l@{}l@{}l} competition: does a\\ public option in-\\crease competition?  \end{tabular} & 
    \begin{tabular}[c]{@{}l@{}l@{}l@{}l@{}l@{}l@{}l@{}l@{}l@{}l@{}l}``the case against: the public plan will unfairly crowd\\ out private coverage''. heritage foundation. july 28 2009:\\``it 's simply impossible to believe the claims by sen.\\ charles schumer (d-n.y.) and others that \colorbox{SkyBlue}{\textbf{congress}}\\ really will do nothing to disrupt the level playing field\\ by favoring the public plan. with congress as both\\ umpire and a team manager one thing is clear: it will\\ favor its own team. the result is the \colorbox{SkyBlue}{\textbf{public}} plan will\\ unfairly crowd out \colorbox{SkyBlue}{\textbf{private}} coverage. \end{tabular} & 
    \begin{tabular}[c]{@{}l@{}l@{}l} congress will not favor pri\\-vate insurer over private\\ insurers\end{tabular} & 
    \begin{tabular}[c]{@{}l@{}l@{}l} \colorbox{SkyBlue}{\textbf{congress}} will not favor \colorbox{SkyBlue}{\textbf{private}}\\ insurer over \colorbox{SkyBlue}{\textbf{public}} insurers  \end{tabular} & 
    \begin{tabular}[c]{@{}l@{}l@{}l} government will favor public\\ insurance; no level playing\\-field \end{tabular} \\
    \bottomrule
    \end{tabular}
}
\label{tab:qualitative}
\vspace{-10pt}
\end{table*}
\begin{table}[t]
\centering
\caption{Effect of the source of constraints to QFS performance on Constrained BART.  $\dag$ denotes significantly better than the other two methods with paired t-test at 0.05 level.}
\resizebox{\columnwidth}{!}{
\begin{tabular}{lccccccc}
\toprule
 \multirow{2}{*}{Constraints} & \multicolumn{3}{c}{\texttt{Debatepedia}} & & \multicolumn{3}{c}{\texttt{PubMedQA}}  \\ \cline{2-4} \cline{6-8}
 & R-1 & R-2 & R-L &  & R-1 & R-2 & R-L \\ \midrule
    Query-only & 61.4 & 48.4 & 59.9 &  & 39.0 & 16.9 & 29.9 \\
    Document-only  & $\textbf{62.9}^\dag$ & $\textbf{50.1}^\dag$ & $\textbf{61.5}\dag$ &  & $\textbf{39.2}$ & $\textbf{17.1}$ & $\textbf{30.1}$ \\
    Query+Document & 61.5 & 48.4 & 60.1 &  & 38.9 & 17.0 & 29.7 \\
    \bottomrule
    \end{tabular}
}
    \label{tab:ablation}
\end{table}

\vspace{1pt}
\noindent
\textbf{\textit{Evaluation Metrics:}} \space
We evaluate the effectiveness of Constrained QFS with ROUGE-1, ROUGE-2, and ROUGE-L \cite{lin2004rouge} for fair comparison to existing works \cite{su2021improve,park2022qsg}.


\noindent
\textbf{\textit{Implementation Details:}} \space
We adopt an off-the-shelf Cross Encoder model\footnote{https://huggingface.co/brutusxu/distilbert-base-cross-encoder-first-p} as our saliency model. We identify the top-3 important tokens with Eq.\ref{eq:integrad} and  construct constraints as $\mathcal{C}=C_1 \land C_2 \land C_3$.

We experiment with two pre-trained Seq2Seq models as the base generator, T5 \cite{raffel2020exploring} and BART \cite{lewis2019bart}. Different from previous works BART-QFS and QSG BART \cite{su2021improve,park2022qsg}, we do not warm start BART or T5 by pre-finetuning on existing abstractive summarization datasets; instead we only finetune them on our target datasets Debatepedia and PubMedQA.
For T5, we format the input as
\texttt{Summarize: Document:} \textit{d} \texttt{Question:} \textit{q}\texttt{: } 
and finetune the model weights on each dataset's training set with golden references. 
At inference time, we use the same input format and finetuned model weights for relevance-constrained generation/generation.
For BART, we format the input as
\texttt{[CLS]} \textit{d} \texttt{[SEP]} \textit{q} \texttt{[EOS]},
where \texttt{[CLS]}, \texttt{[SEP]}, \texttt{[EOS]} are special tokens indicating start, separate and end of sequence, then we finetune and generate text in a similar fashion to T5.
For both models, we finetune with AdamW optimizer \cite{loshchilov2017decoupled}, learning rate $2e-5$ and early stop after no improvements on the dev set for three consecutive epochs. We make our code publicly available at \href{https://github.com/zhichaoxu-shufe/Constrained-QFS}{https://github.com/zhichaoxu-shufe/Constrained-QFS}.



\vspace{2pt}
\section{Results and Analysis}
\label{sec:result}

We address two RQs in this section:

\vspace{2pt}
\noindent
\textbullet \, \textbf{RQ1}: How competitive is the proposed constrained generation method in terms of performance compared to baselines?

\vspace{2pt}
\noindent
\textbullet \, \textbf{RQ2}: How does constrained generation affect QFS performance?

\vspace{3pt}
\noindent
To answer \textbf{RQ1}, 
shown in Table~\ref{tab:results}, we observe that the relevance-constrained methods achieve competitive performance on two datasets. On Debatepedia dataset, Constrained-BART achieves near parity with the current state-of-the-art system and substantially outperforms all other baselines. This result is particularly interesting given the reduced complexity Constrained-BART.
On the PubMedQA dataset, Constrained-BART achieves slightly better performance than QSG BART. A possible explanation for this improved performance might be the length of the documents in PubMedQA, where the relevance-constrained process results in a more consistent snippet.
We therefore conclude that the proposed relevance-constrained generation paradigm can achieve competitive performance without additional parameters or finetuning.

To answer \textbf{RQ2}, we specifically draw a comparison between the proposed methods and their unconstrained baselines, which were finetuned end-to-end and generated QFS without constraints. 
In the second section of Table~\ref{tab:results}, we observe that the proposed constrained generation methods consistently outperform their unconstrained counterparts across different datasets and backbone LMs. 
For instance, on the Debatepedia dataset, Constrained-BART outperforms BART 14.9\%  in R-2.
Therefore, we conclude that by adding carefully constructed constraints into the generation stage, the performance of the QFS task can be significantly improved without modifying the backbone LMs.

\vspace{3pt}
\noindent
\textbf{\textit{Qualitative Analysis:}} \space
We show two examples in Table~\ref{tab:qualitative}. 
In the first example, the BART generation hallucinates "public owners" that is not faithful to the document; however, Constrained-BART is able to successfully summarize the document as $\mathcal{C}$ contains "privatization".
In the second example, despite the underspecified query, the saliency model still extracts critical tokens, which are able to aid in the generation of a  meaningful summary.

\vspace{3pt}
\noindent
\textbf{\textit{Ablation Study:}} \space
In Table~\ref{tab:ablation} we study the effect of different sources of constraints. Query-only denotes that the top-3 important tokens are from the query, and vice versa for Document-only and Query+Document. 
We observe that on the Debatepedia dataset, Document-only constraints significantly outperforms the other two approaches, while on PubMedQA this improvement is minor. 
After manual examination, we find that the golden references in Debatepedia dataset overlap more with documents compared to queries, while PubMedQA does not adhere to this trend. 

\vspace{0pt}
\section{Conclusion and Future Work}
In this work, 
our lightweight relevance-constrained generation approach achieves competitive performance compared to the state-of-the-art method, and it can easily generalize to new domains provided the existence of an effective retrieval model to guide the constraint construction. 
Our future work may involve investigating the effectiveness and summarization faithfulness/factuality of this approach in real world IR systems.

\newpage
\section{Acknowledgement}
Zhichao Xu is supported partially by NSF IIS-2205418 and NSF DMS-2134223. Any opinions, findings and conclusions or recommendations expressed in this material are those of the authors and do not necessarily reflect those of the sponsor.

\bibliographystyle{ACM-Reference-Format}
\balance
\bibliography{reference}

\end{document}